\documentclass[aps,prc,twocolumn,amsmath,amssymb,superscriptaddress,showpacs,showkeys]{revtex4}
\usepackage{dcolumn}
\usepackage{graphicx,color}
\usepackage{textcomp}
\begin {document}
  \newcommand {\nc} {\newcommand}
  \nc {\beq} {\begin{eqnarray}}
  \nc {\eeq} {\nonumber \end{eqnarray}}
  \nc {\eeqn}[1] {\label {#1} \end{eqnarray}}
  \nc {\eol} {\nonumber \\}
  \nc {\eoln}[1] {\label {#1} \\}
  \nc {\ve} [1] {\mbox{\boldmath $#1$}}
  \nc {\ves} [1] {\mbox{\boldmath ${\scriptstyle #1}$}}
  \nc {\mrm} [1] {\mathrm{#1}}
  \nc {\half} {\mbox{$\frac{1}{2}$}}
  \nc {\thal} {\mbox{$\frac{3}{2}$}}
  \nc {\fial} {\mbox{$\frac{5}{2}$}}
  \nc {\la} {\mbox{$\langle$}}
  \nc {\ra} {\mbox{$\rangle$}}
  \nc {\etal} {\emph{et al.}}
  \nc {\eq} [1] {(\ref{#1})}
  \nc {\Eq} [1] {Eq.~(\ref{#1})}
  \nc {\Ref} [1] {Ref.~\cite{#1}}
  \nc {\Refc} [2] {Refs.~\cite[#1]{#2}}
  \nc {\Sec} [1] {Sec.~\ref{#1}}
  \nc {\chap} [1] {Chapter~\ref{#1}}
  \nc {\anx} [1] {Appendix~\ref{#1}}
  \nc {\tbl} [1] {Table~\ref{#1}}
  \nc {\fig} [1] {Fig.~\ref{#1}}
  \nc {\ex} [1] {$^{#1}$}
  \nc {\Sch} {Schr\"odinger }
  \nc {\flim} [2] {\mathop{\longrightarrow}\limits_{{#1}\rightarrow{#2}}}
  \nc {\textdegr}{$^{\circ}$}
  \nc {\inred} [1]{\textcolor{red}{#1}}
  \nc {\inblue} [1]{\textcolor{blue}{#1}}
  \nc {\IR} [1]{\textcolor{red}{#1}}
  \nc {\IB} [1]{\textcolor{blue}{#1}}
\title{Extension of the ratio method to low energy}
\author{F.~Colomer}
\email{frederic.colomer@ulb.ac.be}
\affiliation{Physique Nucl\' eaire et Physique Quantique, Universit\'e Libre de Bruxelles (ULB), B-1050 Brussels}
\author{P.~Capel}
\email{pierre.capel@ulb.ac.be}
\affiliation{Physique Nucl\' eaire et Physique Quantique, Universit\'e Libre de Bruxelles (ULB), B-1050 Brussels}
\author{F.~M.~Nunes}
\email{nunes@nscl.msu.edu}
\affiliation{National Superconducting Cyclotron Laboratory
and Department of Physics and Astronomy,
Michigan State University, East Lansing, Michigan 48824, USA}
\author{R.~C.~Johnson}
\email{r.johnson@surrey.ac.uk}
\affiliation{Department of Physics, University of Surrey,
Guildford GU2 7XH, United Kingdom}
\affiliation{National Superconducting Cyclotron Laboratory
and Department of Physics and Astronomy,
Michigan State University, East Lansing, Michigan 48824, USA}
\date{\today}
\begin{abstract}
{\bf Background:} The ratio method has been proposed as a means to remove the reaction model dependence in the study of halo nuclei. 
{\bf Purpose:} Originally, it was developed for higher energies but given the potential interest in applying the method at lower energy, in this work we explore its validity at 20 MeV/nucleon.
{\bf Method:}  The ratio method takes the ratio of the breakup angular distribution and the summed angular distribution (which includes elastic, inelastic and breakup) and uses this observable to constrain the features of the original halo wave function. In this work we use the Continuum Discretized Coupled Channel method and the Coulomb-corrected Dynamical Eikonal Approximation for the study.
{\bf Results:}  We study the reactions  of $^{11}$Be on $^{12}$C, $^{40}$Ca and $^{208}$Pb at 20 MeV/nucleon. We compare the various theoretical descriptions and explore the dependence of our result on the core-target interaction. 
{\bf Conclusions:}  Our study demonstrates that the ratio method is valid at these lower beam energies.
\end{abstract}
\maketitle
%


\section{Introduction}\label{introduction}
The development of the halo phenomena when approaching the nuclear driplines has become a focus of many studies.
New candidates for halos continue to emerge \cite{nakamura2009, takechi2014,Nak14,Kob14} and specific properties of known halos continue to provide challenges to nuclear theory \cite{whitmore2015}. Detailed reaction studies with halos continue to help us understand the complexity of the reaction mechanism (e.g. Refs.~\cite{schmitt2012,fukui2015}). Given the different energy scales involved in the halo nucleons relative to the excitation energies of the core, effective field theories are now being used to explore halo nuclear structure \cite{ryberg2014,ji2014}. The halo phenomenon is one that connects nuclear physics to other areas, such as atomic and molecular physics, where it can be better controlled through external fields (e.g. \Ref{stipanov2014}).

The continued interest in nuclear halos calls for improved methods  in the extraction of their properties from reaction observables. The most popular way to study halo nuclei is through breakup reactions. Breakup cross sections are large, and they contain information about the binding energy, angular momentum and size of the original halo system \cite{capel2006}. However, the analysis of a breakup experiment contains also uncertainty in the reaction model, particularly the effective interactions used to describe the system. Of special concern is the core-target interaction, which is usually not well known. In Refs.~\cite{capel2011,capel2013} we propose the ratio method that circumvents this ambiguity. There were two works that inspired this new method. 
A recent detailed analysis of the elastic-scattering and breakup cross sections for 
a one-neutron halo showed that the angular distribution for these two processes exhibit very similar diffraction patterns \cite{capel2010}.
This interesting result is easily explained within the Recoil and Excitation Breakup model (REB) developed in \Ref{johnson1997}, in which the angular distributions for both elastic scattering and breakup factorize into a cross section for a pointlike projectile times a form factor that accounts for the extension of the projectile's halo.
The point cross section being identical for both processes explains the strong similarities between the elastic-scattering and breakup cross sections observed in \Ref{capel2010}.
This also means that by taking the ratio of breakup and elastic angular distributions, one can remove most of the dependence on the reaction mechanism and hence obtain a reaction observable sensitive only to the projectile structure.

The original studies on the ratio method \cite{capel2011,capel2013} focused on reactions at around 70 MeV/nucleon, because many of previous breakup experiments had been performed in this energy regime \cite{Nak99,Fuk04,Nak09}. However, it is experimentally very challenging to determine both the elastic and the breakup angular distributions with good precision at 70 MeV/nucleon because the process is very forward focused. At lower energy, a much wider angular range would be available for placing detectors, without having to deal with beam dump issues. With this motivation in mind, the goal of this work is to determine whether the ratio method is valid at around 20~MeV/nucleon (appropriate for facilities such as SPIRAL~2 and FRIB) and to what extent will it still be sensitive to the projectile structure.

In our earlier studies \cite{capel2011,capel2013}, the reaction theory used to describe the processes was the Dynamical Eikonal Approximation \cite{baye2005}. A comparative study \cite{capel2012} showed that, around 70 MeV/nucleon, DEA compares very well with the Continuum Discretized Coupled Channel (CDCC) method \cite{austern1987}. However, it is also shown that the validity of DEA is no longer true at 20~MeV/nucleon, and therefore a better theory is necessary. Studies have shown that most of the limitations of DEA at the lower energies arise from the Coulomb deflection  \cite{capel2012}.
A semiclassical Coulomb correction of the DEA was tested in \Ref{fukui2014} (we will refer to it as CC-DEA) and the results demonstrate that this low-energy correction fixes the problem with the original DEA model and allows an extended use at lower energies in Coulomb-dominated reactions.

Our goal being to test the ratio method on both light and heavy nuclei, we need a reaction model that provides reliable elastic-scattering and breakup cross sections on light targets.
CDCC has been a very successful tool in describing reactions of loosely bound systems over the last few decades.
Accordingly, it is widely used to study reactions with halo nuclei (e.g.\cite{moro2010,moro2012,fukui2015}).
CDCC has been recently tested through a detailed comparison with the exact Faddeev formulation for deuteron collisions on various targets at different energies \cite{upadhyay2012}.
The results of this benchmark show that CDCC perfectly reproduces the Faddeev predictions for elastic scattering of the deuteron but at low energy ($E_d\sim 10$~MeV) and large angles ($\theta\gtrsim 40^\circ$).
The CDCC predictions for deuteron breakup deteriorate at low energy (i.e. at $E_d\sim10$--20~MeV), but are in excellent agreement with Faddeev for both angular and energy distributions from $E_d\sim40$~MeV, which is about the energy range we are interested in.
A major difficulty pointed out in \Ref{upadhyay2012} is the coupling with the transfer channel.
The present study involves a heavier projectile than deuteron.
This leads to a larger amount of absorption within the projectile-target interaction, implying that the coupling to the transfer channel will play a much weaker role, and hence that the CDCC expansion will be much more efficient than for a deuteron projectile.

We will thus rely on both CDCC and CC-DEA to perform the investigation of the ratio method at 20~MeV/nucleon. In our study we will look at the reactions of $^{11}$Be on $^{12}$C, $^{40}$Ca and $^{208}$Pb. For the lighter target, for which the CDCC method converges well, we will present CDCC results along with CC-DEA. The agreement of CDCC and CC-DEA predictions provides reassurance that indeed these methods are valid at these energies. For the heavier targets, for which CDCC exhibits convergence issues, we will rely only on CC-DEA. In all cases, we will compare the computed ratio with the prediction from the REB model to ultimately test the validity of the method at lower energies.

The paper is organized as follows. In \Sec{theory} we introduce our theoretical framework as well as a brief summary of the REB model and the ratio idea. In \Sec{sensitivity} we discuss the sensititivy of the ratio observable to the beam energy and the details of the halo wave function. In \Sec{results} we present our results for the various targets and perform a comparison between the ratios obtained. Finally, in \Sec{conclusion} we draw our conclusions.


\section{Theoretical framework}\label{theory}

To describe the collision of one-neutron halo nuclei on a target, we consider a three-body model of the reaction.
The projectile $P$ of mass $m_P$ is described as a two-body structure: a neutron $n$ of mass $m_n$ loosely bound to a core $c$ of mass $m_c$ ($m_P=m_c+m_n$).
The core is assumed to be in its ground state of spin and parity $0^+$ and its internal structure is neglected. 
To reduce the computational time, the spin of the halo neutron is neglected.
This simplification does not have any effect on the ratio observable.
 The Hamiltonian $H_0$ corresponding to this description reads
\beq
H_0=-\frac{\hbar^2}{2\mu}\Delta_r+V_{cn}(\ve{r}),
\eeqn{e1}
where $\ve{r}$ is the relative coordinate of the halo neutron $n$ to the core $c$, $\mu=m_cm_n/m_P$ is the $c$-$n$ reduced mass, and $V_{cn}$ is a phenomenological potential simulating the $c$-$n$ interaction. The eigenstates of $H_0$ describe the physical states of the projectile
\beq
H_0\ \phi_{lm}(E,\ve{r})=E\ \phi_{lm}(E,\ve{r}),
\eeqn{e2}
where $E$ is the $c$-$n$ relative energy, $l$ is the $c$-$n$ orbital angular momentum and $m$ is its projection.
The negative-energy states ($E<0$) are discrete and correspond to the bound states of the projectile.
The positive-energy states ($E>0$) correspond to the $c$-$n$ continuum, they describe the broken-up projectile.
The parameters of $V_{cn}$ are fitted to reproduce the energies and quantum numbers of the low-lying states of the projectile.

The target $T$ is seen as a structureless particle of mass $m_T$ and charge $Z_T e$, whose interaction with the projectile constituents is simulated by optical potentials $V_{cT}$ and $V_{nT}$.
These potentials, chosen from the literature, reproduce the elastic-scattering cross section of the core and the neutron with the target.

Within this three-body model, describing the collision of the projectile onto the target reduces to solving the three-body \Sch equation
\beq
\left[-\frac{\hbar^2}{2\mu_{PT}}\Delta_R+H_0+V_{cT}(\ve{R}_{cT})+V_{nT}(\ve{R}_{nT})\right]\Psi(\ve{r},\ve{R})\nonumber\\
=E_{\rm tot}\ \Psi(\ve{r},\ve{R}),
\eeqn{e3}
where $\ve{R}$ is the projectile-target relative coordinate, $\mu_{PT}$ is their reduced mass, and $\Psi$ is the three-body wave function.
Within this Jacobi set of coordinates, the $c$-$T$ and $n$-$T$ relative coordinates are
$\ve{R}_{cT}=\ve{R}-\frac{m_n}{m_P}\ve{r}$ and $\ve{R}_{nT}=\ve{R}+\frac{m_c}{m_P}\ve{r}$, respectively.

Equation~\eq{e3} must be solved with the condition that the projectile is initially in its ground state $\phi_{l_0m_0}$ of energy $E_0$.
With $Z$ as the direction of the incoming beam, this condition reads
\beq
\Psi(\ve{r},\ve{R})\flim{Z}{-\infty}e^{i\left\{K_0Z+\eta \ln\left[K_0(R-Z)\right]\right\}}\ \phi_{l_0m_0}(E_0,\ve{r}),
\eeqn{e4}
where $\hbar K_0$ is the $P$-$T$ initial relative momentum, which is related with the total energy by $E_{\rm tot}=\hbar^2K_0^2/2\mu_{PT}+E_0$, and $\eta = Z_T Z_P e^2/(4\pi\epsilon_0\hbar^2K_0/\mu_{PT})$ is the $P$-$T$ Sommerfeld parameter, $Z_P e$ being the charge of the projectile.

In the Continuum Discretized Coupled Channel method (CDCC), the three-body wave function is written in terms of the complete set of eigenstates of the $c$-$n$ system, including bound and continuum states. In our implementation, the continuum is discretized into continuum bins, averaging the scattering states over energy or momentum. Introducing this expansion of the three-body wave function into \Eq{e3} gives rise to the CDCC equations, which are then solved in a truncated model space, with scattering boundary conditions. More details can be found in \Ref{book}.

The Dynamical Eikonal Approximation (DEA) is based on the eikonal approximation \cite{glauber}, which assumes that at sufficiently high energy, the projectile-target relative momentum does not deviate much from the incoming one $\hbar K_0\ve{\hat Z}$.
Following that assumption, the three-body wave function $\Psi$ can be factorized as a plane wave times a function  varying smoothly with $\ve{R}$.
This factorisation leads to a significant simplification of the \Sch equation \eq{e3} \cite{baye2005,goldstein2006}, hence enabling us to perform reaction calculations within a shorter computational time than within a full CDCC framework.
In the following low-energy calculations, we use the Coulomb-corrected version of the DEA (CC-DEA) detailed in \Ref{fukui2014}.

In the Recoil Excitation and Breakup model \cite{johnson1997} \Eq{e3} is solved exactly under two simplifying assumptions: neglecting the $n$-$T$ interaction, i.e.\ assuming $V_{nT}=0$, and considering the adiabatic---or sudden---approximation, i.e. $H_0-E_0\approx 0$.
Under these conditions, Johnson {\it et al.} \cite{johnson1997} prove that the elastic-scattering cross section factorizes into a cross section for a pointlike projectile times a form factor that accounts for the spatial extension of the $c$-$n$ wave function.
This same factorization occurs for the inelastic and break-up cross sections as shown in \Ref{RCJ1997}, page 160, Eq.~(18), and exploited in Refs~\cite{capel2011,capel2013}.

In Refs.~\cite{capel2011,capel2013} we introduce then the ratio observable:
\beq
{\cal R}_{\rm sum}(E,\ve Q) &=& \frac{(d\sigma/dEd\Omega)_{\rm bu}}{(d\sigma/d\Omega)_{\rm sum}}\label{ratio-sum},
\eeqn{reb-sum}
where the summed cross section corresponds to
\beq
\lefteqn{\left(\frac{d\sigma}{d\Omega}\right)_{\rm sum}}\nonumber\\
 &=&\left(\frac{d\sigma}{d\Omega}\right)_{\rm el}
+\sum_{i>0}\left(\frac{d\sigma_i}{d\Omega}\right)_{\rm inel}+\int \left(\frac{d\sigma}{dEd\Omega}\right)_{\rm bu} dE.
\eeqn{xs-sum}

As described in Refs.~\cite{capel2011,capel2013}, if the REB model is  valid, then this ratio of cross sections should correspond exactly to the form factor connecting the halo ground state and the continuum state with relative energy $E$ populated in the breakup
\beq
{\cal R}_{\rm sum}(E,\ve Q) & \stackrel{\rm (REB)}{=} & |F_{E,0}(\ve Q)|^2\;.
\eeqn{r-ff}
The form factor introduced reflects the structure of the projectile:
\beq
\lefteqn{|F_{E,0}(\ve Q)|^2 =}\nonumber\\
&&\frac{1}{2l_0+1} \sum_{m_0}\sum_{lm} \left| \int\phi_{lm}(E,\ve{r}) \phi_{l_0m_0}(E_0,\ve{r})
e^{i\ve{Q\cdot r}}d\ve{r}\right|^2,
\eeqn{reb-ff2}
and depends on  $\ve Q = \frac{m_n}{m_P} (K_0\ve{\widehat{Z}} - \ve K')$ which is proportional to the momentum transferred
during the scattering process. 

\section{Sensitivity of the ratio observable}\label{sensitivity}

\subsection{Dependence on beam energy}\label{beam}
\begin{figure}[t]
\includegraphics[width=8cm]{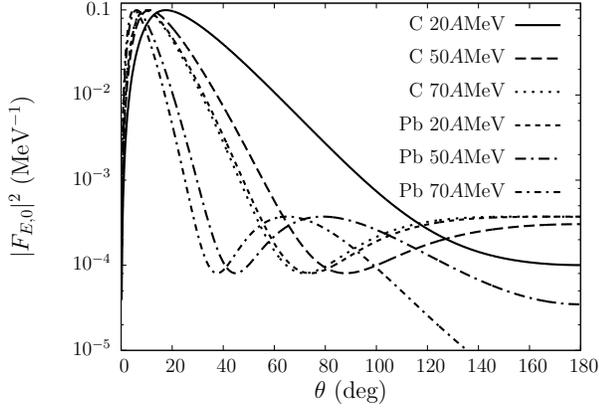}
\caption{Ratio form factor as a function of different beam energies and different targets. Results for a carbon target are shown by the solid line (at 20~MeV), the dashed line (at 50~MeV) and the dotted line (at 70~MeV). The other lines are for the lead target. }
\label{f-ratio1}
\end{figure}

In order to understand what happens to the form factor when changing the beam energy or the target, we investigate \Eq{reb-ff2}. The dependence of  $|F_{E,0}(\ve Q)|^2$ on the beam energy or the target mass appears only through $\ve Q$. 
The momentum transfer modulates the diffraction pattern contained in the point-like cross section and dictates  the relevant scattering angles to be considered in the process
\beq
Q=2\frac{m_n}{m_P} K_0 \sin(\theta/2).
\eeqn{eQ}
For a given target, the larger the beam energy, the larger $K_0$ and therefore increasing the beam energy squeezes the distribution to smaller angles.
Identically, for a given beam energy, if one increases the target mass,  $K_0$ increases, producing a similar effect. This is illustrated in Fig.\ref{f-ratio1}  

\subsection{Sensitivity to the projectile structure}\label{structure}

To evaluate the information that can be obtained from the ratio observable at 20~MeV/nucleon for a $^{12}$C target, we plot a figure similar to the Fig.~1 of \Ref{capel2011}.
As expected from our previous analysis, we observe a significant dependence on the projectile description.
The ratio varies in both shape and magnitude when the binding energy is changed.
It is also strongly dependent on the angular momentum of the bound state.
However, as explained above, a lower energy and target mass leads to a ratio that extends over a larger angular range, more favorable to an experimental use of the method.
\begin{figure}[h]
\includegraphics[width=8cm]{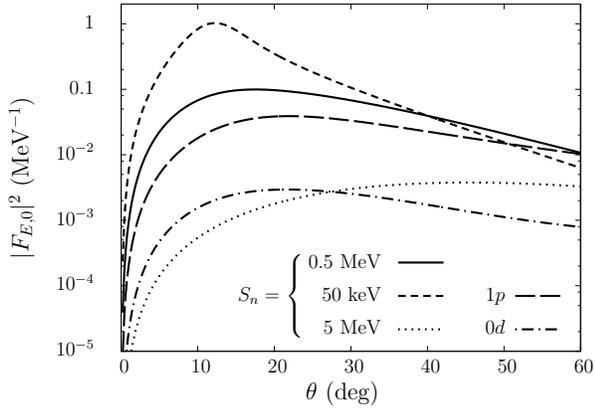}
\caption{Sensitivity of the form factor to the binding energy and the partial wave of the halo neutron to the core: included at the realistic $^{11}$Be, a $1s$ state bound by 0.5 MeV (solid line), a loosely bound $1s$ state $S_n=0.05$ MeV (short-dashed line) and a well bound $1s$ state $S_n=5$ MeV (dotted line). Also shown are the ratio for a $1p$ state with $S_n=0.5$ MeV (long-dashed line) and for a $0d$ state with $S_n=0.5$ MeV (dash-dotted line) }
\label{f4}
\end{figure}

\begin{figure}[h]
\includegraphics[width=8cm]{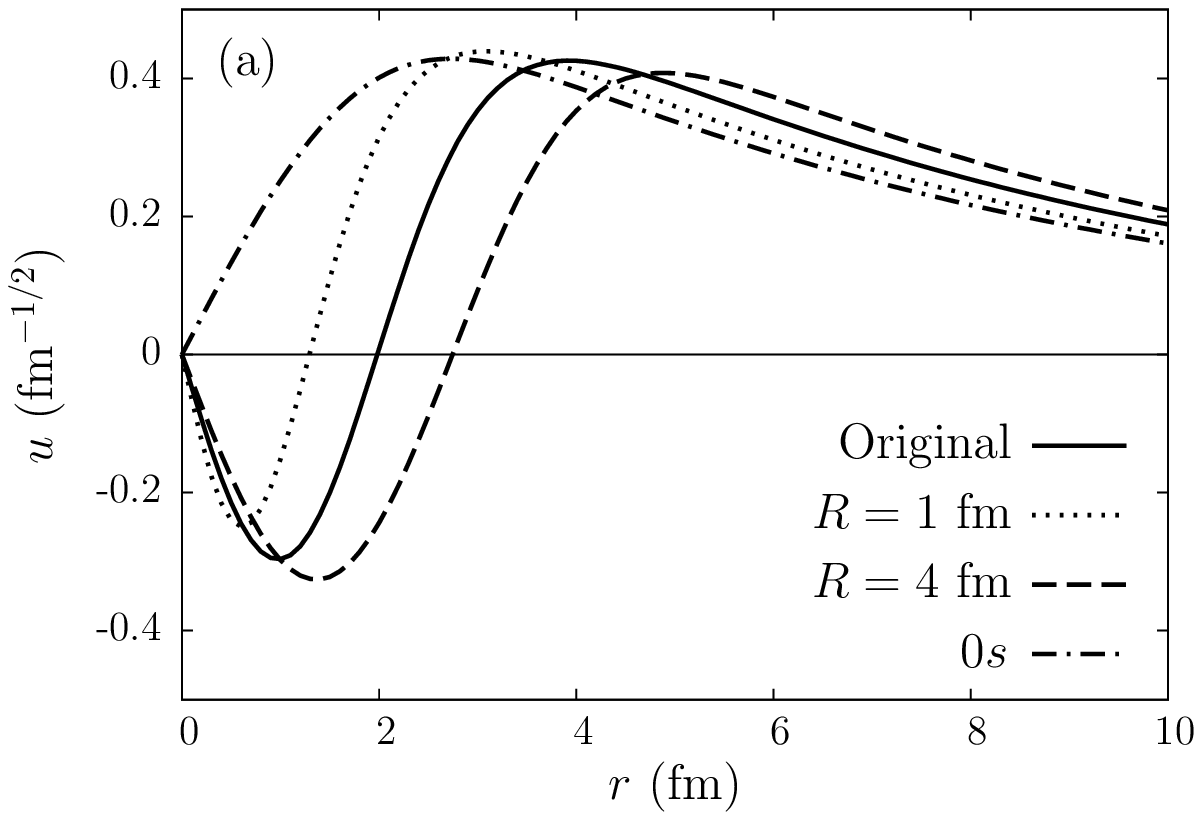}\\
\includegraphics[width=8cm]{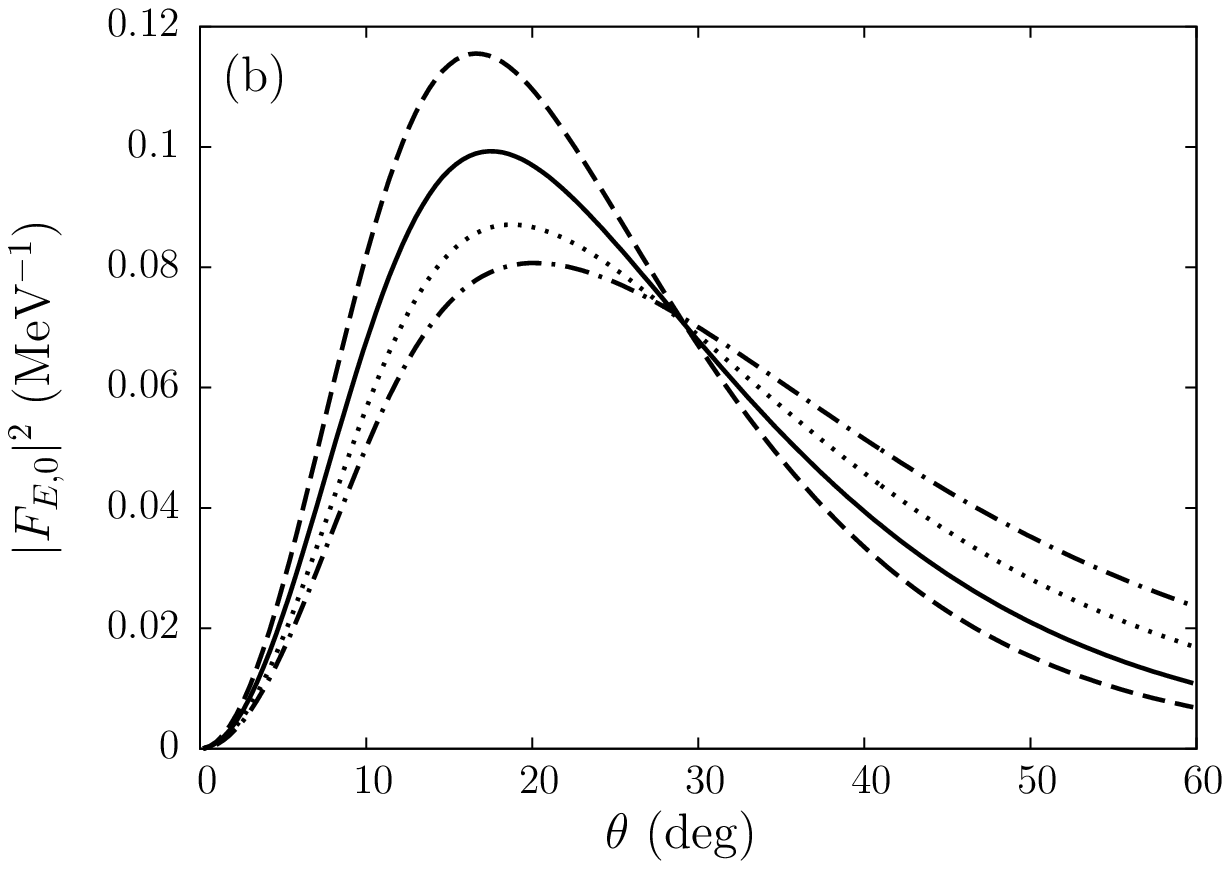}\\
\includegraphics[width=8cm]{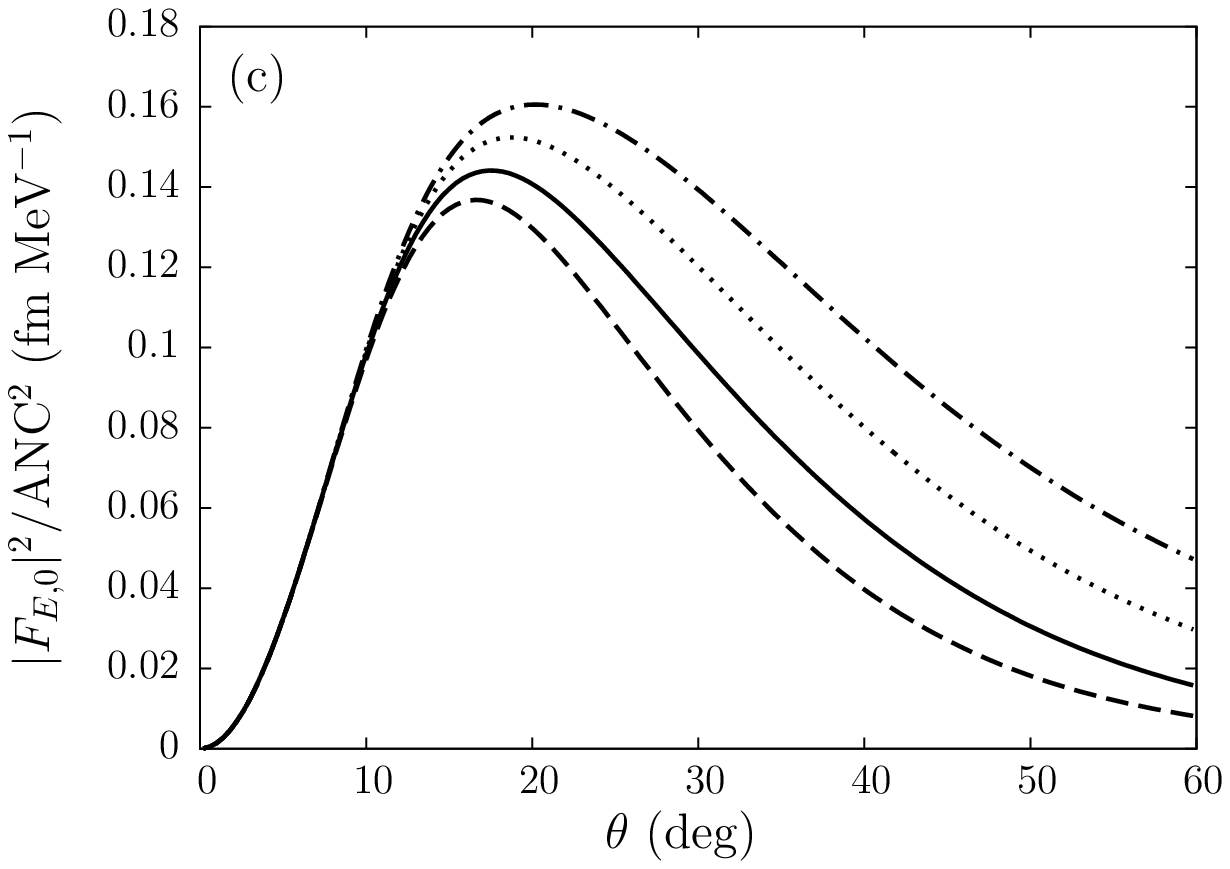}
\caption{Sensitivity of the form factor to the radial wave function: (a) the different reduced radial wave function considered in this test: the original $^{11}$Be wave function (solid line), for a small radius (dotted line) and a large radius (dashed line). Also included is the results for a $0s$ state (dash-dotted line), to show the effect of the node in the wave function.
(b) the corresponding form factors \eq{reb-ff2}.
(c) the form factor divided by the square of the ANC of the ground state wave function $\phi_{l_0 m_0}$.}
\label{f5}
\end{figure}

We have also looked at the sensitivity to the radial wave function and have observed a similar result as in \Ref{capel2013}.
Using different geometries of the $c$-$n$ potential, we have generated a set of bound-state radial wave functions that vary both in their internal and external parts [\fig{f5}(a)].
Although less sensitive to this part of the wave function, the ratio provides a clean test of the projectile radial wave function, unlike most of the other reaction observables  [see \fig{f5}(b)].
As already observed, it probes different parts of the wave function depending on the scattering angle \cite{capel2013}.
At forward angle, the ratio scales perfectly with the square of the asymptotic normalisation constant (ANC) of the wave function, as shown in \fig{f5}(c), where the form factor \eq{reb-ff2} has been scaled by the square of the ground-state ANC.
At larger angles, this is no longer the case indicating that it becomes sensitive to the internal part of the wave function.

\section{Reaction calculations}\label{results}

\subsection{Numerical details}\label{numerics}

As mentioned in the Introduction, to test the validity of the Ratio Method at low beam energy, we compare the REB prediction \eq{r-ff} to dynamical calculations within the CDCC and CC-DEA frameworks, which
are considered to provide the exact ratio.
In this section, we provide the numerical details of these calculations.

We first define the effective interactions used in constructing our three-body Hamiltonian. For the $^{10}$Be-$n$ interaction, we take the same parameters as those of \Ref{capel2013}, but neglect the spin of the halo neutron for simplicity.
As to the optical potentials, we use the Koning-Delaroche global parameterization for the neutron-target interaction \cite{KD}.
For the $^{10}$Be-target interactions, we adapt the $^{12}$C-target potentials from \Ref{sahm1986}, simply scaling their radius to account for the mass of the projectile.
The mathematical expression and parameters of these optical potential are given in Appendix~\ref{potApp}.

CDCC calculations are performed with the code {\sc fresco} \cite{fresco}. We were only able to obtain fully converged breakup cross sections at these lower beam energies for the $^{12}$C target. In this case,  the model space is defined as follows:  $^{10}$Be-$n$ partial waves up to $l_{\rm max} = 6$ and $Q_{\rm max} = 6$ multipoles in the expansion of the coupling potentials. The coupled equations are integrated up to $R_{\rm max} = 60$ fm, and the scattering wave functions are matched at $R_{\rm asym}=1000$ fm. Cross sections include  up to total angular momentum $J_{\rm max} = 20000$.

The CC-DEA calculations are performed following \Ref{goldstein2006} including the semiclassical Coulomb correction detailed in \Ref{fukui2014}.
The DEA equation is solved with the algorithm presented in \Ref{capel2003}, in which the projectile wave-function is expanded over a spherical mesh.
Lowering the projectile energy requires an increase of the number of points on the unit sphere.
At 20~MeV/nucleon, 
it has to go up to $N_\theta\times N_\phi=16\times 31$ for the C target, 
 $N_\theta\times N_\phi=14\times 27$ 
 for the Ca target,
and $N_\theta\times N_\phi=12\times 23$ 
is sufficient for the Pb target.
The quasi-uniform radial grid contains 600 points (800 for the Pb target) and extends up to 600~fm (800~fm for the Pb target).
The calculations are performed for impact parameters $b=0$--100~fm (C target) or 160~fm (Ca and Pb targets) with a discretization step that varies between 0.25~fm and 5~fm.
As explained in \Ref{goldstein2006}, the angular distributions are obtained with an extrapolation up to $b_{\rm max}=200$~fm (C target) or 300~fm (Ca and Pb targets).


\subsection{Carbon target}\label{carbon}
\begin{figure}[h]
\includegraphics[width=8cm]{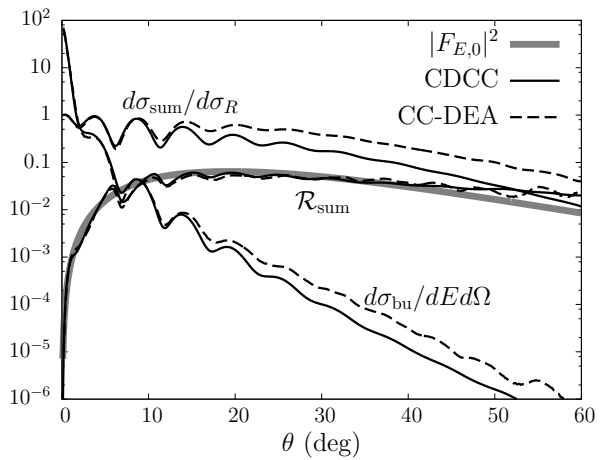}
\caption{Comparison between CDCC (solid black lines) and CC-DEA (dashed lines) calculations for a $^{11}$Be projectile impinging on a C target at 20~MeV/nucleon.
Also shown is the REB form factor \eq{reb-ff2} (thick grey line). 
We consider the breakup to a final core-$n$ scattering state of  $E=125$ keV.}
\label{f1}
\end{figure}

We begin our analysis of the ratio method at low energy considering a carbon target.
At such an energy and for such a target, the CDCC method is the most reliable model on the market.
However, because CC-DEA is much more cost effective than CDCC, we will use this case to test the validity of that model in a nuclear-dominated collision.
The comparison between CDCC (solid line) and CC-DEA (dashed line) is illustrated in \fig{f1}.
This figure displays the summed cross section \eq{xs-sum} (ratio to Rutherford) and
the differential breakup cross section as a function of the center-of-mass scattering angle for a $^{10}$Be-$n$ continuum energy $E=125$~keV (in b/MeV sr).
The rapid drop of the latter indicates that practical measurements could probably be made up to $40^\circ$.
Hence we have limited our study a bit beyond that angle.

We observe that at forward angles, where the reaction is dominated by the Coulomb interaction, CC-DEA is in perfect agreement with CDCC, thanks to the semiclassical correction.
At larger angles, i.e.\ beyond $15^{\circ}$, CC-DEA deviates from CDCC.
At these angles, the reaction is fully dominated by the far side of the $T$ matrix, i.e. by the attractive part of the projectile-target interaction \cite{carlson1985}, and the  Coulomb correction alone is no longer sufficient to obtain reliable individual cross sections.
As already observed in \Ref{alkhalili1997}, the discrepancy between CC-DEA and CDCC are due to non-eikonal effects which increase at low energy.
Nevertheless, it is interesting to note that even though the correct individual angular distributions are not reproduced by the CC-DEA, their ratio \eq{reb-sum} is in agreement with CDCC.
This is probably to be related to the strong independence of the ratio to the reaction process, and hence to the details of the model used to describe it.
We plan to investigate this promising fact in a later work. In any case,
this result indicates that  CC-DEA can indeed be used to test the validity of the REB prediction \eq{r-ff}. 
Since CC-DEA is much more cost-effective than CDCC, we will use it to study the ratio method for the heavier targets (Ca and Pb), for which the (repulsive) Coulomb force between the projectile and the target plays a more dominant role (see Secs.~\ref{Ca} and \ref{Pb}).

\begin{figure}[h]
\includegraphics[width=8cm]{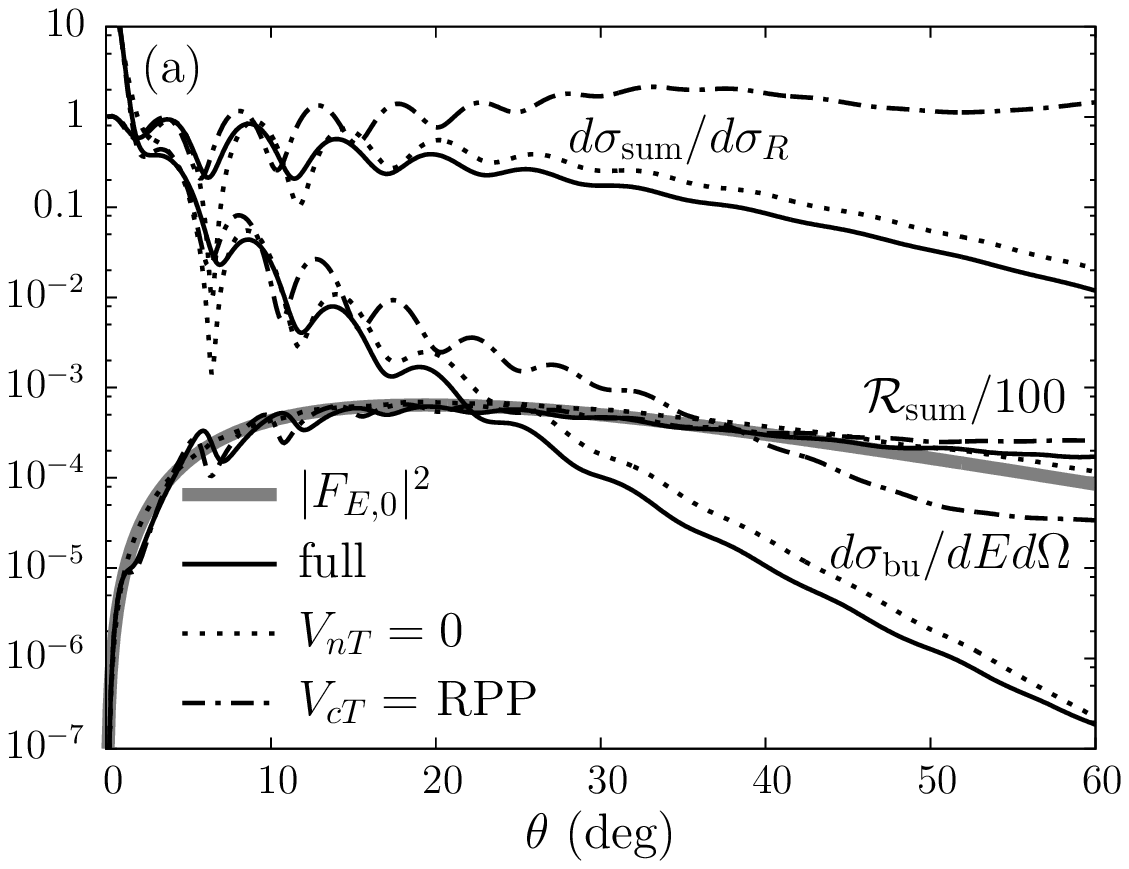}\\
\includegraphics[width=8cm]{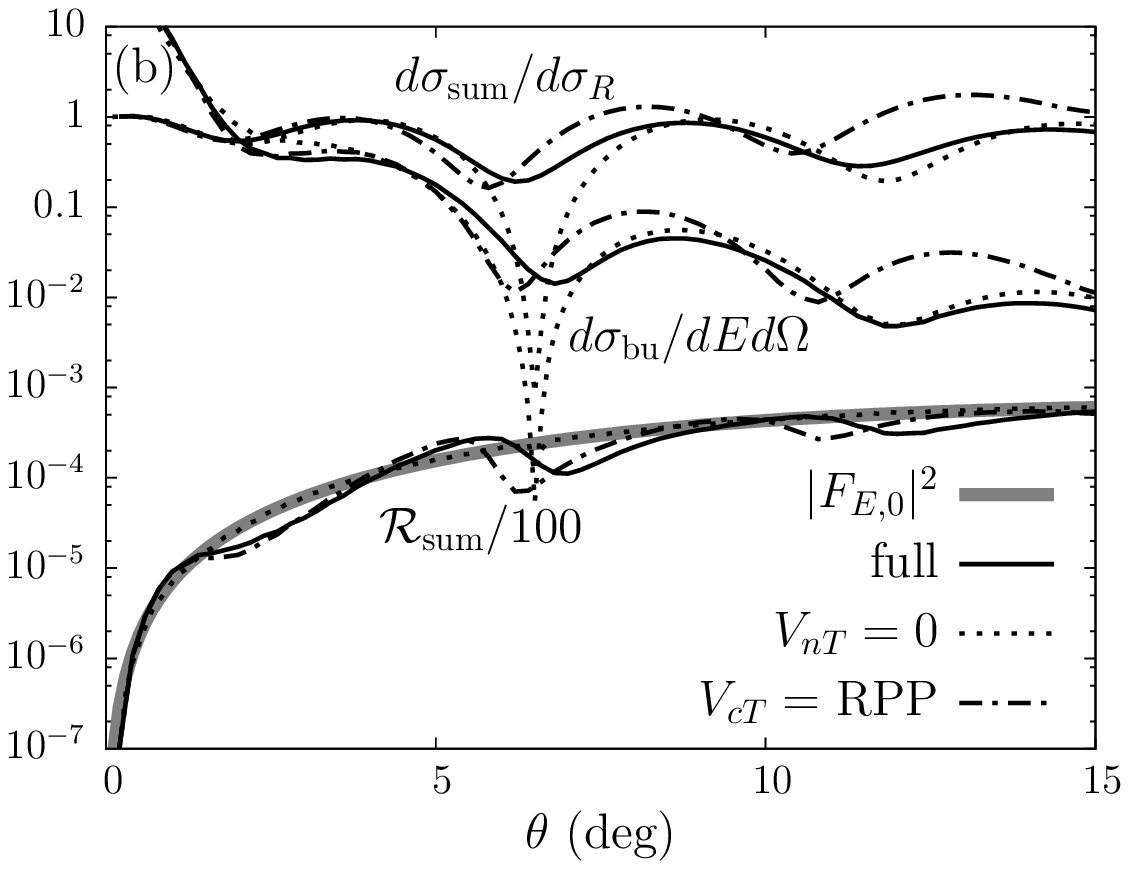}
\caption{Sensitivity of the ratio to the $P$-$T$ interaction for a C target at 20MeV/nucleon. We consider the breakup to a final core-$n$ scattering state of  $E=125$ keV.
The original calculation (solid lines) is compared to the results obtained with $V_{nT}=0$ (dotted lines) and when $V_{cT}=$~RPP (dash-dotted lines).
The REB form factor \eq{reb-ff2} is plotted as well (thick grey line).
(a) full angular range (using CDCC).
(b) forward-angle region (using CC-DEA).}
\label{f2}
\end{figure}

In \fig{f2}, we analyse the sensitivity of the ratio method to the $P$-$T$ interaction. 
The upper panel shows the sensitivity of the ratio in a wide angular range using CDCC, while
the lower panel zooms in on the forward angle region (CC-DEA calculations).
The solid lines correspond to the full calculation, which includes both $c$-$T$ and $n$-$T$ interactions as mentioned in \Sec{numerics}.
As already observed at higher energy \cite{capel2010,capel2011,capel2013}, the breakup and summed cross sections exhibit very similar features: same oscillatory pattern at forward angles and similar decay at larger angles.
Accordingly, their ratio is very smooth, confirming that this reaction observable removes most of the sensitivity to the reaction process (note that the ratio has been divided by 100 for readability); we merely observe remnant oscillations in the $5^\circ$--$15^\circ$ range [see \fig{f2}(b)].
In addition, the actual ratio follows very closely the REB prediction \eq{reb-ff2} (thick grey line), indicating that information about the projectile structure can be reliably extracted from the ratio as emphasized in \Sec{structure}.

Besides that full calculation, \fig{f2} displays the results obtained including only the $c$-$T$ interaction (i.e. setting $V_{nT}=0$, dotted lines).
As already observed before \cite{capel2011,capel2013}, switching off that interaction leads to a near-perfect agreement with the REB prediction: the dotted-line ratio is nearly superimposed to the form factor $|F_{E,0}|^2$.
This confirms the role played by the $n$-$T$ interaction, which is to shift slightly the angular distribution \cite{johnson1997}, leading to the small remnant oscillations in the actual ratio between $5^\circ$ and $15^\circ$.

More interesting is the result obtained changing $V_{cT}$ to another potential.
Following \Ref{capel2004}, we have considered a potential developed to reproduce the elastic scattering of $^{10}$B on a carbon target at 18~MeV \cite{robson1971,perey1976} (RPP, dash-dotted lines).
The large difference with the potential of  \Ref{sahm1986} enables us to test the (in)sensitivity of the ratio to that interaction.
As already seen in \Ref{capel2004}, this second potential leads to a complete change in the cross section: not only is the oscillatory pattern shifted in angle, but for both elastic scattering and breakup, the cross sections are significantly increased.
However, the ratio remains nearly unchanged.
This spectacular result confirms that even at low energy, the ratio method enables us to remove most of the dependence on the $P$-$T$ interaction, leading to an observable uniquely sensitive to the projectile structure.

\subsection{Ca target}\label{Ca}
\begin{figure}[h]
\includegraphics[width=8cm]{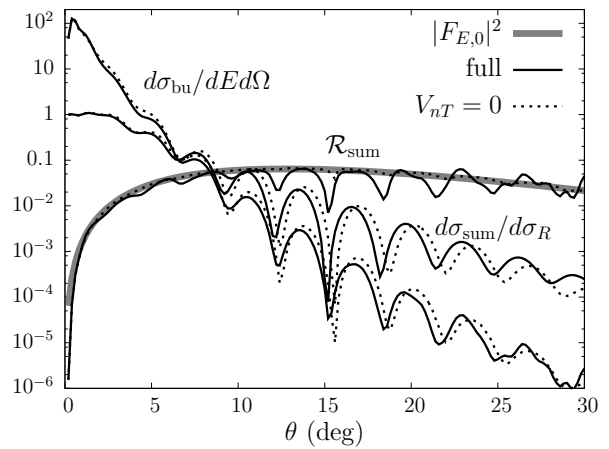}
\caption{Ratio prediction for $^{11}$Be impinging on a Ca target at 20MeV/nucleon (CC-DEA calculations). We consider the breakup to a final core-$n$ scattering state of  $E=125$ keV.}
\label{f3}
\end{figure}

To study the target dependence of the ratio, we perform a series of calculations for a calcium target at 20~MeV/nucleon.
For this target, we could not obtain converged CDCC cross sections.
Fortunately, the angular range of interest is limited to the region of near-far interferences, up to which the CC-DEA is reliable as shown in the previous section.

The results are summarized in \fig{f3}.
Compared to the carbon target, we observe significant changes in the angular distributions.
The summed cross section shows that at forward angles the collision is dominated by Rutherford scattering, due to the higher $Z$ of the target.
We also note that both angular distributions fall off more rapidly with the scattering angle $\theta$ than for the carbon target, meaning that the breakup cross section becomes very small beyond $25^\circ$.
The oscillatory pattern due to the near-far interference is more pronounced and extends over a larger angular range than for the carbon target.
Yet, as observed in \Ref{capel2010}, both the summed and breakup cross sections exhibit similar features, as predicted by the REB.
Considering their ratio \eq{reb-sum} removes most of these angular dependences leading to a reaction observable in excellent agreement with the REB prediction \eq{r-ff}.
As observed at higher energies \cite{capel2011,capel2013} and on the carbon target, the remnant oscillations caused by the $n$-$T$ interaction disappear when $V_{nT}=0$  (dotted lines).

These results confirm the validity of the ratio method at low energy and its independence of the target choice and of the reaction process.

\subsection{Lead target}\label{Pb}

\begin{figure}[h]
\includegraphics[width=8cm]{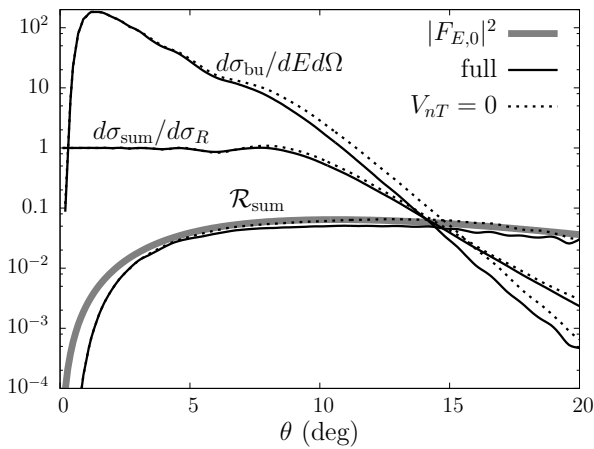}
\caption{Same as Fig. \ref{f3} but for a Pb target.}
\label{f4}
\end{figure}
We have also performed calculations for a Pb target at 20~MeV/nucleon.
For such a heavy target, we could not obtain convergence of the summed and breakup cross sections in CDCC.
While the rate of convergence of the ratio was better than the individual angular distributions, these were also not converged, although they were approaching the REB prediction.
Instead we have used the CC-DEA, 
which proved to work very well at this energy \cite{fukui2014}.
In CC-DEA, we were only able to make predictions out to $20^\circ$ due to numerical instabilities.  Results from CC-DEA for Pb are not so promising as those for C and Ca.
The CC-DEA predictions for the ratio follows the same trend as the REB prediction, but remains away from it.
At the most forward angles, the adiabatic approximation leads to an underestimation of the REB ratio.
As already observed in \Ref{capel2013}, this underestimation worsens at low beam energy.
Hopefully, a non-adiabatic correction, such as the one developed by Summers \etal\ in \Ref{summers2002}, could improve the REB prediction of the ratio for heavy targets at low energies.
We plan to study such a correction in the near future.
At larger angles, the $n$-$T$ interaction keeps the DEA result away from the form factor as already discussed in \Ref{capel2013}.

These results show that although the REB predicts a target independent ratio, it is best to use light targets due to the breakdown from the REB prediction for the heaviest systems. This comes mostly from the adiabatic approximation at these low energies, which works better for nuclear-dominated reactions.
Moreover, as shown in \Sec{sensitivity}, the use of light targets helps spreading the range of the ratio form factor towards larger angles, better suited for an experimental application of the method.

\section{Conclusions}\label{conclusion}
The ratio \eq{reb-sum} is a new reaction observable suggested to study the structure of loosely-bound quantal structures, such as halo nuclei.
It is predicted to be nearly independent of the reaction mechanism while capturing the projectile structure.
In this work we test the validity of the method at low energies, given the potential interest of applying the method at facilities such as SPIRAL~2 and FRIB.
We have performed CDCC and CC-DEA calculations for $^{11}$Be impinging on C, Ca and Pb at 20 MeV/nucleon, to obtain elastic and breakup cross sections, from which the ratio is obtained.
We have then compared the results of the calculations with the REB prediction for the form factor $|F_{E,0}|^2$ to determine the validity of the method.

Our results show that the ratio method is valid at lower energy and thus can be used in a larger number of facilities. The fact that the form factor is spread over a larger angular range makes it easier for the setup of the experiment. At the same time one needs to be aware of the magnitude of the breakup cross section when moving toward the larger angles, as it becomes very low.
We expect measurements with $^{11}$Be beams on carbon targets to be possible out to at least $40$ degrees. Although breakup cross sections are larger for the heaviest targets, the method works best for lighter targets because the adiabatic approximation assumed within the REB breaks down in Coulomb-dominated processes at low energy.
A non-adiabatic correction to the REB form factor may solve that problem \cite{summers2002}.
According to these results, it seems very likely that the ratio method could be extended to even lighter targets, and in particular to protons.
Experimentally, elastic and inelastic scattering off a proton target can be measured by the missing mass method.
Albeit promising, this possibility would require testing within a reliable reaction model valid for such light targets, viz. the Faddeev-AGS theory \cite{deltuva2005}.

We have also demonstrated that the sensitivity to the projectile description is equally present in the ratio observables extracted from these lower energy reactions. Indeed, although measurements at these energies may allow a larger angular range, the sensitivity to the internal part of the wave function is also pushed out to larger angles.

Another important result coming from the present study is the realization that even though at larger angles the CC-DEA cross sections do not reproduce the CDCC results, the ratio predicted by CC-DEA is in complete agreement with CDCC.
This suggests that even a simple dynamical description of the reaction may provide an accurate ratio. This is another appealing motivation to use the ratio method: thanks to its independence of the reaction mechanism it may not require the use of state-of-the-art reaction models to analyse accurately experimental data.
We plan to study this aspect of the ratio method in a future analysis within the framework of perturbative approaches.

This work opens the ratio method to a larger number of facilities, particularly those with lower beam energies, and will motivate groups to collect data for an experimental test of the method.

\section*{Acknowledgments} 
This work was supported by the National Science
Foundation under Grant No.~PHY-1403906, the Department
of Energy, Office of Science, Office of Nuclear Physics under award No.~DE-FG52-08NA28552, and the Research Credit No.~19526092 of the Belgian Funds for Scientific Research F.R.S.-FNRS.
RCJ is supported by the UK STFC through Grant No. ST/F012012/1.
This text presents research results of the Belgian Research Initiative on eXotic nuclei
(BriX), program No.~P7/12 on inter-university attraction poles of the Belgian Federal Science Policy Office.

\appendix
\section{Two-body interactions}\label{potApp}

This appendix details the form factors of the optical potentials and the parameters used in this study to describe the core-$n$, the core-target and $n$-target interactions. 

As presented in Sec. \ref{theory}, the projectile is described as a two-body structure. The $V_{cn}$ potential contains only a central real volume term
\beq
V_{cn}(\ve{r}) = -V_rf(r,R_r,a_r),
\eeqn{eApp1}
with the Woods-Saxon form factor
\beq
f(r,R_r,a_r)=\left[1+\exp\left(\frac{r-R_r}{a_r}\right)\right]^{-1}
\eeqn{eApp2}
of radius $R_r$ and diffuseness $a_r$.
The parameters of the core-$n$ potentials are listed in \tbl{TabPotGS}.
We take the same parameters as those of \Ref{capel2013}, but neglect the spin of the halo neutron for simplicity. This potential reproduces the experimental binding energy 0.5~MeV in the $1s$ orbital to describe the $1/2^+$ ground state of $^{11}$Be.
The same $V_{cn}$ potential is used to obtain the continuum wave functions appearing in \Eq{reb-ff2}.

\begin{table}[t]
\caption{\label{TabPotGS} Parameters of the core-$n$ potential.}

\centering
\begin{tabular}{l c c c}
  \hline
  \hline                      
 $V_r$& $R_r$ & $a_r$ \\
 (MeV)& (fm) & (fm) \\
     \hline
62.52 & 2.585 & 0.6 \\
  \hline
  \hline
\end{tabular}
\end{table}

The nuclear part of the $V_{xT}$ potential that simulates the interaction between the projectile fragment $x$ ($c$ or $n$) and the target contains both real and imaginary volume terms as well as an imaginary surface term :
\beq
V_{xT} &=& -V_rf(r,R_r,a_r) -i W_rf(r,R_i,a_i) \nonumber\\
	&&-i W_D\,a_D\frac{d}{dr}f(r,R_D,a_D).
\eeqn{eApp3}
The Coulomb part of $V_{cT}$ is simulated by the potential due to a uniformly charged sphere of radius $R_C$. 

We use the Koning-Delaroche global parameterization for the neutron-target interaction \cite{KD}.
For the $^{10}$Be-target interactions, we adapt the $^{12}$C-target potentials from \Ref{sahm1986}, by simply scaling their radius to account for the mass of the projectile.
In \fig{f5} we also display results obtained using a potential developed by Robson \cite{robson1971} and listed in the Perey and Perey compilation \cite{perey1976}.
The parameters used in this study are listed in \tbl{TabPot}.

\begin{table*}[t]
\caption{\label{TabPot} Parameters of the optical potentials simulating the interaction between the fragments of the projectile and the target nucleus.}

\centering
\makebox[\columnwidth]{\begin{tabular}{c c c c c c c c c c c c c}
  \hline
  \hline                      
  $P$ & $T$ & Ref. & $V_r$ & $R_r$ & $a_r$ & $W_i$ & $R_{i}$ & $a_{i}$ & $W_D$& $R_{D}$ & $a_{D}$ & $R_C$ \\
   & & & (MeV) & (fm) & (fm) & (MeV) & (fm) & (fm) & (MeV) & (fm) & (fm) & (fm) \\
  \hline
 $n$ & $^{12}$C  & \cite{KD} & 46.9395 & 2.5798 & 0.676 & 1.8256 & 2.5798 & 0.676 & 28.6339 & 2.9903 & 0.5426 & $-$ \\
  	& $^{40}$Ca & \cite{KD} & 46.709 & 4.054 & 0.672 & 1.728 & 4.054 & 0.672 & 28.926 & 4.406 & 0.538 & $-$ \\
	 & $^{208}$Pb & \cite{KD} & 41.4872 & 7.3202 & 0.6469 & 1.1858 & 7.3202 & 0.6469 & 26.4580 & 7.3973 & 0.5102 & $-$ \\
 $^{10}$Be & $^{12}$C & \cite{sahm1986} & 250 & 3.053 & 0.788 & 247.9 & 2.982 & 0.709 & 0 & $-$ & $-$ & 5.777 \\
        & $^{12}$C &  \cite{robson1971,perey1976} & 100 & 5.40 & 0.5 & 18 & 5.40 & 0.5 & 0 & $-$ & $-$ & 5.40 \\
  	& $^{40}$Ca & \cite{sahm1986} & 200 & 4.465 & 0.837 & 276.9 & 5 & 0.653 & 0 & $-$ & $-$ & 4.465 \\
	 & $^{208}$Pb & \cite{sahm1986} & 95 & 7.0129 & 1.168 & 250 & 7.9582 & 0.662 & 0 & $-$ & $-$ & 10.503\\
  \hline
  \hline
\end{tabular}}
\end{table*}


\end{document}